\documentclass{ptephy_v1}

\preprintnumber{XXXX-XXXX} 

\begin{document}

\title{Efficiency at maximum power output for an engine with a passive piston}


\author{Tomohiko G. Sano\footnote{Current address: Department of Physics, Ritsumeikan University, Kusatsu, 525-8577 Shiga, Japan.\\ E-mail: tomohiko@gst.ritsumei.ac.jp} and Hisao Hayakawa}
\affil{Yukawa Institute for Theoretical Physics, Kyoto University Kitashirakawa Oiwakecho, Sakyo-ku, Kyoto 606-8502 Japan \email{tomohiko@yukawa.kyoto-u.ac.jp}}


\begin{abstract}%
Efficiency at maximum power (MP) output for an engine with a passive piston without mechanical controls between two reservoirs is theoretically studied. We enclose a hard core gas partitioned by a massive piston in a temperature-controlled container and analyze the efficiency at MP under a heating and cooling protocol without controlling the pressure acting on the piston from outside. We find the following three results: (i) The efficiency at MP for a dilute gas is close to the Chambadal-Novikov-Curzon-Ahlborn (CNCA) efficiency if we can ignore the side wall friction and the loss of energy between a gas particle and the piston, while (ii) the efficiency for a moderately dense gas becomes smaller than the CNCA efficiency even when the temperature difference of reservoirs is small. (iii) Introducing the Onsager matrix for an engine with a passive piston, we verify that the tight coupling condition for the matrix of the dilute gas is satisfied, while that of the moderately dense gas is not satisfied because of the inevitable heat leak. We confirm the validity of these results using the molecular dynamics simulation and introducing an effective mean-field-like model which we call stochastic mean field model.
\end{abstract}

\subjectindex{xxxx, xxx}

\maketitle

\section{Introduction}
Equilibrium thermodynamics reveals the relation between work and heat, and the upper bound for extracted work from an arbitrarily heat cycle \cite{calen,landau}. 
The milestone of equilibrium thermodynamics is that thermodynamic efficiency for any heat cycle between two reservoirs characterized by the temperatures $T_{\rm H}$ and $T_{\rm L}$ $(T_{\rm H} > T_{\rm L})$ is bounded by the Carnot efficiency: $\eta_{\rm C} \equiv 1 - T_{\rm L}/T_{\rm H}$ achieved by quasi-static operation \cite{carnot}. 
There are many studies on the efficiency of engines including both external and internal combustion engines. The steam engines and steam turbines belong to the former category whose ideal cycles are the Carnot cycle, the Stirling cycle and so on \cite{carnot, stirling}. The diesel and free-piston engines are examples of the latter, and their ideal cycles are the Otto cycle, the Brayton cycle and so on \cite{internal, internal_book}. It is also known that the maximum efficiency for the ideal external combustion engines is $\eta_{\rm C}$, while that for the ideal internal ones is usually smaller than $\eta_{\rm C}$.
For a practical point of view, an engine with $\eta_{\rm C}$ is useless, because its power is zero.

The extension of thermodynamics toward finite-time operations, so-called finite time thermodynamics, has been investigated by many authors \cite{chambadal, novikov, CA,vdb,esposito_2010,izumida,seifert_2011,CV2011,hop,izumida_onsager,izumida_nonlinear,otto1,otto2,otto3,otto4,otto5,otto6,ftt_text, engine_nature,schmiedl_prl,schmiedl_epl, Reitlinger, yvoon,Vaudrey_2014,zu_2008,rana_2014}. 
Chambadal and Novikov independently proposed, and later Curzon and Ahlborn rediscovered that the efficiency at maximum power output (MP) is given by the Chambadal-Novikov-Curzon-Ahlborn (CNCA) efficiency: $\eta_{\rm CA} \equiv 1 - \sqrt{T_{\rm L}/T_{\rm H}}$ \cite{chambadal, novikov, yvoon,CA,Reitlinger, Vaudrey_2014}. 
Recently it is found that Reitlinger originally proposed $\eta_{\rm CA}$ in 1929 \cite{Reitlinger, Vaudrey_2014}. 
The validity of the CNCA efficiency near equilibrium has been justified through the linear irreversible thermodynamics \cite{vdb}, molecular kinetics \cite{izumida,izumida_onsager} or low-dissipation assumption \cite{esposito_2010}. 
It is believed that the CNCA efficiency is, in general, only the efficiency at MP near equilibrium situations. Indeed, there are many situations to exceed the CNCA efficiency in idealized setups \cite{esposito_2010, hop,izumida}. 
Although there are several studies for finite time thermodynamics including external and internal combustion engines or fluctuating heat engines \cite{engine_nature,schmiedl_prl,schmiedl_epl,zu_2008,rana_2014}, they are mostly interested in force-controlled engines \cite{izumida,seifert_2011,CV2011,izumida_onsager,izumida_nonlinear,otto3,otto4,otto5,otto6,hop,engine_nature,schmiedl_prl,schmiedl_epl, rana_2014}, where a piston or a partitioning potential is controlled by an external agent. On the other hand, the efficiency at MP for an engine partitioned by a passive piston without any external force control, has not been well-studied so far. 

\begin{figure}[h]
\begin{center}
\includegraphics[scale =0.6]{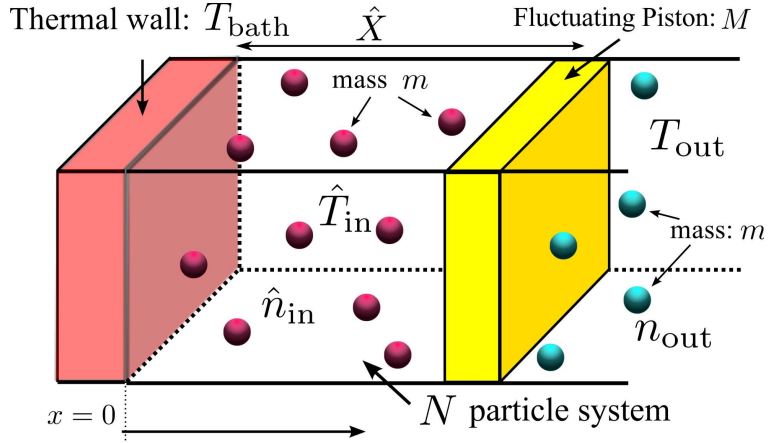}
\caption{(Color online). A schematic picture of our setup, where $N$ identical hard core particles are enclosed in a container partitioned by an adiabatic piston of mass $M$ at $x = \hat{X}$. The density $n_{\rm out}$ and the temperature $T_{\rm out}$ for the outside gas $x > \hat{X}$ are kept to be constants. The temperature $T_{\rm bath}$ of the thermal wall at $x = 0$ is controlled by an external agent, while thermodynamic quantities such as the density $\hat{n}_{\rm in}$ and the temperature $\hat{T}_{\rm in}$ fluctuate in time.}
\label{setup}
\end{center}
\end{figure}

The aim of this paper is to clarify the efficiency at MP for the engine with a passive piston, which is an idealized model of internal combustion engines without mechanical controls. We consider a hard core gas confined by a massive piston in a chamber, where the piston freely moves in one-direction by the pressure difference (see Fig. \ref{setup}). We use the molecular dynamics (MD) simulation of hard core gases to examine a theoretically derived efficiency at MP on the basis of an effective model, which we call stochastic mean field model (SMF).

Because the engine we consider is an internal combustion engine, the maximum efficiency is smaller than the Carnot efficiency. 
Our study is relevant from the following two reasons. 
Firstly, we can find many situations, where the direct mechanical control of a piston is difficult. 
For example, the structure of internal combustion engines is usually too complicated to control inside mechanically \cite{internal_book}. Therefore, we need to clarify the effect of the uncontrollable motion of a piston on the efficiency.
Secondly, the study of engines having passive pistons is important even for finite time thermodynamics. 
In the absence of mechanical control of a piston or a partitioning wall, heat flow when we attach a thermal wall is inevitable. 
Because heat flow from a reservoir is not usually taken into account in conventional finite time thermodynamics, it is important to verify whether the existing theoretical results are unchanged under the existence of such heat flow \cite{chambadal, novikov, CA,vdb,esposito_2010,izumida,seifert_2011,CV2011,hop,izumida_onsager,izumida_nonlinear,otto1,otto2,otto3,otto4,otto5,otto6,ftt_text, schmiedl_prl,schmiedl_epl}. 
Indeed, we will show that conventional results are only valid for our system when the heat flow is negligible as in dilute gases.
Thus, we believe that our study for the simplest engine with a passive piston from a thermodynamic point of view is important.

The organization of this paper is as follows. We explain our setup and operation protocol for the temperature of the thermal wall $T_{\rm bath}$ in Sec. \ref{sec_setup}. 
We introduce SMF in Sec. \ref{effective_model_sec} to analyze the power and efficiency. 
We examine the validity of SMF in Sec. \ref{time_ev} comparing the time evolution of MD simulation and that of SMF.
In Sec. \ref{sec_exmp}, we obtain the efficiency at MP for our engines containing dilute hard core gases theoretically, which is close to the CNCA efficiency in the massive piston limit. We also find that the efficiency at MP for moderately dense gases is smaller than the CNCA efficiency even in the linear non-equilibrium regime. 
In Sec. \ref{sec_ons}, to clarify the efficiency in the linear non-equilibrium regime, we explicitly derive the Onsager matrix. 
We clarify the finite density effect for the efficiency and stress the importance of the heat flux when we attach a bath at $T_{\rm bath}$ on the efficiency at MP in this section. 
We discuss the difference between our results and previous results in Sec. \ref{diss_sec} and conclude the paper with some remarks in Sec. \ref{sec_con}. In App. \ref{derive_qwall}, we show part of the derivation of SMF. In App. \ref{grad_temp_md}, we discuss the time evolution of the temperature profile after we attach a hot reservoir. 
In App. \ref{def_work}, the definition of the work and heat for our system is discussed. In App. \ref{mass_inel}, the effects of piston mass and inelasticity of the piston are studied and we discuss the effect of the sidewall-friction on the piston in App. \ref{friction_effect}. Throughout this paper, variables with ``$\ \hat{}\ $" denote stochastic variables.

\section{Setup}\label{sec_setup}

In our system, $N$ hard core particles of each mass $m$ and diameter $d_{\rm in}$ are enclosed in a three-dimensional container partitioned by an adiabatic piston of mass $M$ and the area $A$ on the right side of $x$-direction, a diathermal wall attached with a thermal bath on the left side of $x$-direction and four adiabatic walls on the other directions (Fig.\ 1). There exists a constant pressure satisfying $P_{\rm out} = n_{\rm out}T_{\rm out}$ from out side of the piston (right side of the piston). The density $n_{\rm out}$ and the temperature $T_{\rm out}$ for the outside gas $x > {X}$ are kept to be constants. 
We assume that adhesion between particles and the walls of the container as well as the one between particles can be ignored. 
The piston is assumed to move in one dimension without any sidewall friction.
Post-collision velocity $(v',V')$ and pre-collision velocity $(v,V)$ in $x$-direction for a colliding particle and the piston are related as:
\begin{eqnarray}
v'(v,V) &=& v -\frac{1}{m}P_v\label{rule1},\\
V'(v,V)  &=& V + \frac{1}{M}P_v\label{rule2},
\end{eqnarray}
where the contribution from the horizontal motion of particles to the wall is canceled as a result of statistical average. Here, $P_v = P_v(V) \equiv M({V}' -{V}) = (1+e) mM (v-{V})/(m+M)$ represents the momentum change of the piston because of the collision for the particle of velocity $v$, where $e$ is the restitution coefficient between the particles and the piston. The reason why we introduce the restitution coefficient is that the wall consists of a macroscopic number of particles and part of impulses of each collision can be absorbed into the wall as the excitation of internal oscillation.

We adopt the Maxwell reflection rule for a collision between a particle and a diathermal wall attached with the bath at $T_{\rm bath}$. The post-collisional velocity ${\boldmath v}' = (v_x ',v_y ', v_z')$ toward the wall at $x = 0$ is chosen as a random variable obeying the distribution
\begin{equation}
\phi_{\rm wall}({\boldmath v}', T_{\rm bath}) = \frac{1}{2\pi}\left(\frac{m}{ T_{\rm bath}}\right)^2v_x ' \exp\left[-\frac{m{\boldmath v'} ^2}{2T_{\rm bath}}\right],
\end{equation}
whose domain is given by $0<v_x '< \infty$ and $-\infty<v_y ',v_z '<\infty$.

\begin{figure}[h]
\begin{center}
\includegraphics[scale = 0.5]{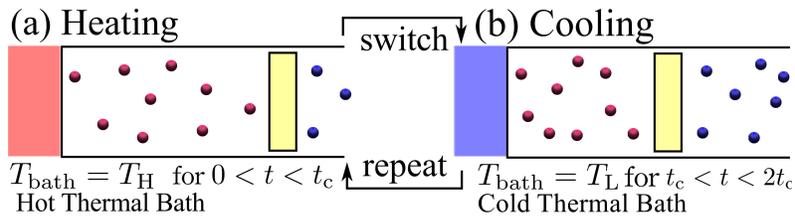}
\caption{(Color online) A set of schematic figures of the operation protocol. We attach a heat bath at $T_{\rm H}$ on the diathermal wall at $t = 0$. (a) For $0<t<t_{\rm c}$, $T_{\rm bath}$ is kept to be $T_{\rm bath} = T_{\rm H}$,  and at $t = t_{\rm c}$, $T_{\rm bath}$ is switched to be that at $T_{\rm L}$ simultaneously, and (b) $T_{\rm bath}$ is kept to be this state until $t = 2t_{\rm c}$. Then, we again replace the bath by that at $T_{\rm H}$ simultaneously. After repeating the switching of $T_{\rm bath}$, the heat cycle reaches a steady cycle.}
\label{protocol}
\end{center}
\end{figure}

Let us consider a heat cycle for heating $T_{\rm bath} = T_{\rm H} > T_{\rm out}$ and cooling  $T_{\rm bath} = T_{\rm L} = T_{\rm out}$ processes (Fig. \ref{protocol}). Initially, the enclosed gas and the gas outside are in a mechanical equilibrium state, which satisfies $\hat{P}_{\rm in} = P_{\rm out}$ and $\hat{T}_{\rm in} = T_{\rm out} = T_{\rm bath}$. At $t = 0$, we attach a heat bath at $T_{\rm H}$ on the diathermal wall. For $0<t<t_{\rm c}$, $T_{\rm bath}$ is kept to be $T_{\rm H}$ (Fig. \ref{protocol} (a)), and at $t = t_{\rm c}$, $T_{\rm bath}$ is switched to be $T_{\rm L}$ simultaneously, and is kept to be this state until $t = 2t_{\rm c}$. Then, we again replace the bath at $T_{\rm L}$ by the one at $T_{\rm H}$ simultaneously (Fig. \ref{protocol} (b)). After repeating the switching and attaching of the baths, the heat cycle reaches a steady cycle. It should be noted that the enclosed gas is no longer thermal equilibrium during the cycle. During the operation, we ignore the time necessary for the switching the heat bath. The finite switching time only lowers the power but does not affect the efficiency of the cycle and what the maximum-power-output process is. 

In this paragraph, we explain some additional remarks for the MD simulation.
We assume that particles are colliding elastically each other and with side walls. 
The collision rule between the piston and a particle is given by Eqs. (\ref{rule1}) and (\ref{rule2}). We introduce typical length and time scale as $X_{\rm ini} \equiv N T_{\rm out}/P_{\rm out}A$ and $t_0 \equiv X_{\rm ini}\sqrt{M/T_{\rm out}}$ for later convenience. The number of particle $N = 200$ is fixed through our simulation. The collisional force from outside the piston is modeled by $\hat{F}_{\rm out}$ as will be defined in Eq. (\ref{force_gas}).

\section{Stochastic mean field model}\label{effective_model_sec}
Let us introduce the stochastic mean field (SMF) model to describe the dynamics of the piston and the energy balance of our system by using two independent stochastic variables: fluctuating density $\hat{n}_{\rm in}(t) = N/A\hat{X}(t)$ and fluctuating temperature $\hat{T}_{\rm in}(t)$. The reason why we call our model the SMF is that the piston moves in a stochastic manner because of impulses of the hard-core particles and we average out the spatial inhomogeneity of the gas. Here, $\hat{n}_{\rm in}(t)$ and $\hat{V}\equiv d\hat{X}/dt$ satisfy the stochastic equations:
\begin{eqnarray}
\frac{d\hat{n}_{\rm in}}{dt} &=& -\frac{\hat{n}_{\rm in}}{\hat{X}}\hat{V},\\
M\frac{d\hat{V}}{dt} &=& \hat{F}_{\rm in} + \hat{F}_{\rm out},\label{eom}
\end{eqnarray}
where the stochastic force $\hat{F}_{\nu} (\nu = {\rm in, out})$ is introduced as
\begin{eqnarray}
\hat{F}_{\nu} &\equiv& \sum_v P_v \cdot \hat{\xi}_{\nu} ^{v}(t|\hat{V},\hat{n}_{\nu}, \hat{T}_{\nu})\label{force_gas}.
\end{eqnarray}
Here, $\hat{\xi}^v _{\rm in}$ and $\hat{\xi}^v _{\rm out}$ denote Poissonian noises of the unit amplitude whose event probabilities are respectively given by 
\begin{eqnarray}
\hat{{\lambda}}_{\rm in} ^{v} &\equiv&dv |v-\hat{V}|\Theta(v-\hat{V})  \hat{n}_{\rm in} \phi_0(v,  \hat{T}_{\rm in})\left\{1 + 4\hat{\Phi} g_0(\hat{\Phi})\right\}\label{lambda_mod},\\
\hat{\lambda}_{\rm out} ^v &\equiv& dv |v-\hat{V}|\Theta(\hat{V}-v) {n}_{\rm out} \phi_0(v,  {T}_{\rm out}),\label{lambda_def}
\end{eqnarray}
where we have introduced the radial distribution function at contact $g_0$ \cite{torquato}. The symbol $``\cdot"$ in Eq. (\ref{force_gas}) represents It\^o type stochastic product \cite{gardiner,kampen,sano_hayakawa}. $\Theta(x)$ is Heaviside function satisfying $\Theta(x) = 1$ for $x \geq 0$ and $\Theta(x) = 0$ for $x  < 0$. The density and temperature for the gas outside are kept to be constants in time, i.e., $\hat{n}_{\rm out} \equiv n_{\rm out}$ and $\hat{T}_{\rm out} \equiv T_{\rm out}$. We introduced the velocity distribution function (VDF) for the gas as $\phi_0(v,\hat{T}_{\rm in}) \equiv \sqrt{{m}/{2\pi  \hat{T}_{\rm in}}}\exp\left[-{mv^2}/{2 \hat{T}_{\rm in}}\right]$. It should be noted that a set of Eqs. (\ref{eom}) and (\ref{force_gas}) is an extension of our previous study toward a finite density hard core gas when the density and the temperature change in time and this is the reason why we adopt It\^o product in Eq. (\ref{force_gas}) \cite{sano_hayakawa}. 
We adopt the equation of state for hard core gases of volume fraction $\hat{\Phi} \equiv \hat{n}_{\rm in}\pi d^3 /6$ is given by \cite{resibor}
\begin{eqnarray}
\hat{P}_{\rm in} = \hat{n}_{\rm in}\hat{T}_{\rm in}(1 + 4\hat{\Phi} g_0(\hat{\Phi}))\label{eos_hard}.
\end{eqnarray}

Next, we propose the time evolution for $\hat{T}_{\rm in}$. The differential of the internal energy for the gas $\hat{U}_{\rm in} \equiv 3N \hat{T}_{\rm in}/2$ is given by
\begin{eqnarray}
d\hat{U}_{\rm in} &=& d\hat{Q}_{\rm wall} + d\hat{E}_{\rm pis}\label{energy_cons_modify},\\
d\hat{Q}_{\rm wall} &\equiv& {d\hat{Q}_{0}} + d\hat{Q}_{J},\label{q_wall_pwall1}\\
\frac{d\hat{Q}_{0}}{dt}&\equiv& A\hat{n}_{\rm in} (T_{\rm bath} - \hat{T}_{\rm in})\sqrt{\frac{2 \hat{T}_{\rm in}}{\pi m}}\label{qwall},\\
\frac{d\hat{E}_{\rm pis}}{dt} &\equiv& \sum_v \frac{m}{2}\left\{{v'}^{2}(v,\hat{V}) -{v}^2\right\} \cdot \hat{\xi}_{\rm in} ^v(\hat{V},\hat{n}_{\rm in}, \hat{T}_{\rm in}),\\
d\hat{Q}_{J} &=&-\frac{45\sqrt{\pi}}{64}\hat{J}_{\rm in}Adt.\label{qwall1}
\end{eqnarray}
Here, $d\hat{Q}_{\rm wall}$ denotes the total heat flow from the thermal bath at $T_{\rm bath}$. $d\hat{Q}_{J}$ denote the heat flux from the internal thermal conduction $\hat{J}_{\rm in}$ and $d\hat{Q}_0$ represents the remaining heat flow $d\hat{Q}_0 = d\hat{Q}_{\rm wall} - d\hat{Q}_{J}$  \cite{izumida,izumida_onsager}. $d\hat{E}_{\rm pis}$ denotes the kinetic energy transfer from the piston to the gas. In summary, main part of our SMF model consists of two coupled equations: the equation of motion for the piston (\ref{eom}) and the energy equation for the enclosed gas (\ref{energy_cons_modify}). In App. \ref{derive_qwall}, we derive
Eqs. (\ref{q_wall_pwall1}), (\ref{qwall}), and (\ref{qwall1}).

\begin{figure}[h]
\begin{center}
\includegraphics[scale = 0.6]{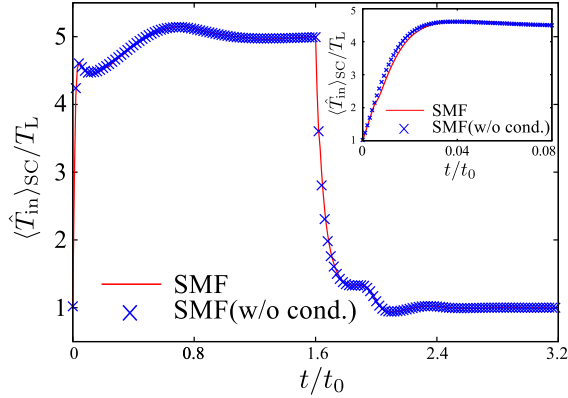}
\caption{(Color online) Comparison between SMF and the SMF without heat conduction (w/o cond.) for $d_{\rm in}/\sqrt{A} = 0.01$ and $t_c/t_0 = 1.60$. The initial volume fraction is calculated to be $\Phi = 1.05\times10^{-4}$. The solid line and the cross points represent the dynamics of temperature for SMF and the dilute version of SMF, respectively. The inset represents the detailed time evolution for $0<t/t_0<0.08$.}
\label{temp_smf_dilute}
\end{center}
\end{figure}

The heat flux $J_{\rm in}$ is estimated from the solution of the heat diffusion equation for the temperature profile ${\mathcal T} = {\mathcal T}(x,t)$:
\begin{equation}
\frac{\partial {\mathcal T}}{\partial t} - \frac{\kappa}{n}\frac{\partial^2 {\mathcal T}}{\partial x^2} = 0\label{heat_eq},
\end{equation}
under the situation that the thermal conductivity $\kappa$ and density $\hat{n}_{\rm in} = n$ are constants in space and time, where the piston position is fixed at $\hat{X} = L$. Imposing the boundary conditions ${\mathcal T}(x,t=0) \equiv {T}_{\rm ini}$, ${\mathcal T}(x=0,t) = T_{\rm bath}$ and $\partial_x {\mathcal T}(x=L,t) = 0$ on Eq. (\ref{heat_eq}), the solution of Eq. (\ref{heat_eq}) is given by
\begin{eqnarray}
{\mathcal T}(x,t) &=& T_{\rm bath} - (T_{\rm bath} - T_{\rm ini})\sum_{l=1} ^{\infty} \frac{4}{\pi l}e^{-\left(\frac{l \pi}{2L}\right)^2\frac{\kappa t}{n}}\sin\left(\frac{l \pi x}{2L}\right)\label{diff_sol}.
\end{eqnarray}
Assuming that $T_{\rm ini},L,\kappa$ and $n$ change in time adiabatically, i.e. $T_{\rm ini}\to \hat{T}_{\rm in}(t), L\to \hat{X}(t), \kappa \to \hat{\kappa}(\hat{\Phi}(t), \hat{T}_{\rm in}(t))$ \cite{resibor,garzo} and $n\to \hat{n}_{\rm in}(t)$, we obtain the approximate heat flux $J_{\rm in} = \int_0 ^{L} - \kappa \partial_x {\mathcal T}(x,t)dx /L$ as
\begin{eqnarray}
\hat{J}_{\rm in}(t) &=& \frac{4\hat{\kappa}}{\pi\hat{X}(t)}(T_{\rm bath} - \hat{T}_{\rm in}(t))\sum_{l=1} ^{\infty}\frac{\sin(l\pi/2)}{l}\exp\left[-\left(\frac{l\pi}{2\hat{X}(t)}\right)^2\frac{\hat{\kappa} t}{\hat{n}_{\rm in}}\right]\label{jin_modify},
\end{eqnarray}
where we have adopted expressions in Refs. \cite{resibor,garzo} for density and temperature dependence of the thermal conductivity $\hat{\kappa}(\Phi, T) = (75\sqrt{T/\pi}/64m d_{\rm in}^2 g_0(\Phi))[\{1+(12\Phi g_0(\Phi)/5)\}^2 + (4608\Phi^2 g_0(\Phi)/225\pi)]$.

Because the heat conduction relaxes fast to a steady state for the dilute gas, we can simplify Eq. (\ref{energy_cons_modify}) as
\begin{eqnarray}
d\hat{U}_{\rm in} &=& d\hat{Q}_{0} + d\hat{E}_{\rm pis}\label{energy_cons},
\end{eqnarray}
though heat conduction exists. We numerically confirm that the gradient of the temperature for the dilute gas relaxes faster than that for the dense gas in App. \ref{grad_temp_md}. Indeed, we compare the dynamics of temperature in Fig. \ref{temp_smf_dilute} for SMF and the SMF without heat conduction using Eq. (\ref{energy_cons}), the difference between two methods is negligible. Here we have adopted the initial volume fraction as $\Phi = 1.05\times10^{-4}$. We choose $t_c/t_0 = 1.60$ which is long enough for the relaxation of the system. We will also show that $d{Q}_{J}$ does not affect the efficiency at MP for the dilute gas later. Thus, we use Eq. (\ref{energy_cons}) for the dilute gas instead of Eq. (\ref{energy_cons_modify}).

In this paragraph, let us explain the numerical details of SMF. The numerical integration is performed through Adams-Bashforth method, with $dt/t_0 \equiv 0.01\epsilon$ and $\epsilon \equiv \sqrt{m/M}$. Calculating $\hat{\xi}_{\nu} ^v$, $v$ and $dv$ are respectively replaced by $v_i$ and $\Delta v$, where $v_i = i \Delta v - v_{\rm max} (i=1,2,\cdots, 600)$, $v_{\rm max} \equiv 6.0 \sqrt{k_{\rm B}T_{\nu}/M} (\nu= {\rm in, out})$ and $\Delta v \equiv v_{\rm max}/ 300$. Because Eq. (\ref{energy_cons_modify}) turns out to be unstable if the heat conduction in Eq. (\ref{qwall1}) is larger than that of Eq. (\ref{qwall}), we impose the condition $d{Q}_{J} = 0$ if $d{Q}_{J} > dQ_{0}$ through the numerical stability of our simulation. The simulation data are averaged in steady cycles, where the averaged quantity is represented by $\langle\cdots\rangle_{\rm SC}$. 

\section{Time evolution}\label{time_ev}
To verify the validity of the SMF model, we compare the time evolution of the MD simulation and SMF. We examine the dilute and moderately dense gases in Sec. \ref{dilute_time_ev} and \ref{dense_time_ev}, respectively.

\subsection{Dilute case}\label{dilute_time_ev}
\begin{figure*}[!t]
\centering
\includegraphics[scale = 0.55]{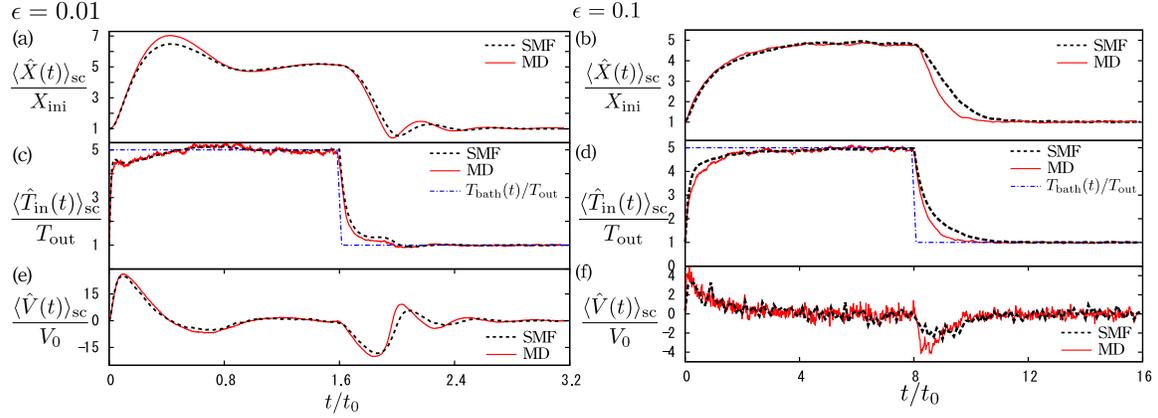}
\caption{(Color online) The time evolutions of steady cycles for $T_{\rm H}/T_{\rm L} = 5.0$. They are categorized into two types: damped oscillating type for $\epsilon = 0.01$ (left) and over-damped type for $\epsilon = 0.1$ (right). The time evolutions for the piston position ((a) and (b)), the temperature ((c) and (d)), and the piston velocity ((e) and (f)) are plotted. Time evolutions for the corresponding physical quantities for MD simulation (solid line) agree with those for the SMF model (dashed line).}
\label{evo}
\end{figure*}

We consider a dilute gas of the diameter $d_{\rm in}/\sqrt{A} = 0.01$ which corresponds to $\Phi = 1.05\times10^{-4}$ at $t = 0$.
Time evolutions of the volume (the position of the piston) for $T_{\rm H}/T_{\rm L} = 5.0$ are drawn in Fig. \ref{evo} (a) for $\epsilon = 0.01$, $t_{\rm c}/t_0 = 1.60$ and (b) for $\epsilon = 0.1$, $t_{\rm c}/t_0 = 8.0$. We have confirmed that this $t_c$ for each $\epsilon$ is larger than the relaxation time to the corresponding steady state.
The simulation data are averaged from 11th cycle to 20th cycle, where the solid and dashed lines, respectively, represent the data for MD simulation and those for simulation of our SMF model. 
Similarly, Figs. \ref{evo} (c) and (d) are the time evolutions for the temperature of the gas, and Figs. \ref{evo} (e) and (f) are the time evolutions for the piston velocity. 
Dot-dashed lines represent the operation protocol of $T_{\rm bath}$. It is remarkable that our SMF model correctly predicts the time evolution of MD.

Let us explain the behavior of the system shown in Fig. \ref{evo}. When the heating process starts, the enclosed gas starts expanding, to find a new mechanical equilibrium density determined by the condition $\hat{P}_{\rm in} = P_{\rm out}$, because the pressure for the enclosed gas becomes larger than that for the outside after the heating. Similarly, the gas is compressed when the cooling process starts. It should be stressed that the heating (cooling) and expansion (compression) processes take place simultaneously. 

\begin{figure}[h]
\begin{center}
\includegraphics[scale = 0.8]{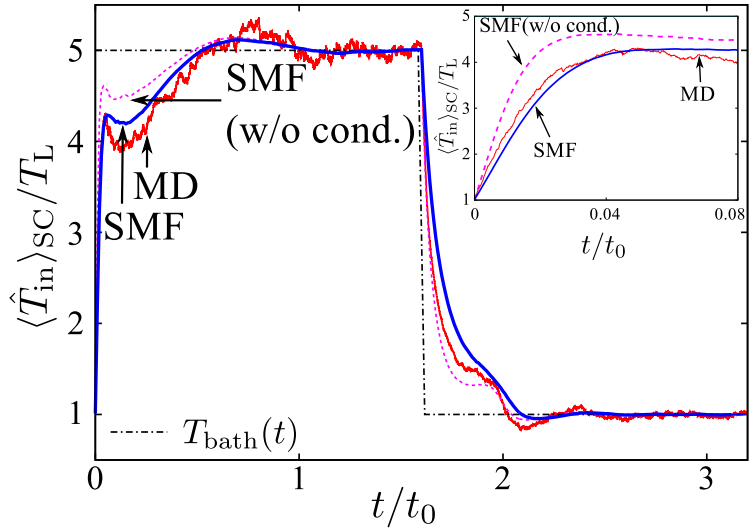}
\caption{(Color online) The time evolution of temperature for MD, the SMF, and the dilute approximation of SMF are compared. For heating regime $t/t_0 < 1.6$, the dilute SMF overestimates the heat gain, while the SMF works better than the dilute version, in particular, for small $t/t_0$. The inset represents the detailed time evolution for $0<t/t_0<0.08$, where SMF captures the MD simulation results.}
\label{time_ev_temp}
\end{center}
\end{figure}

The time evolutions of the physical quantities can be categorized into two types: (a) damped-oscillating type and (b) over-damped type depending on the mass ratio $\epsilon \equiv \sqrt{m/M}$. 
Taking the average of Eq. (\ref{energy_cons}) and assuming that the piston is heavy $\epsilon \ll 1$, the time evolution of the averaged temperature is written as
\begin{eqnarray}
T_{\rm in}(t) &=& T_{\rm bath} \left(1- a_0V(t)\right) + O(\epsilon^2)\label{temp_vs_vel},\\
a_0 &\equiv&\sqrt{\frac{\pi m}{2T_{\rm bath}}} = \epsilon\sqrt{\frac{\pi M}{2T_{\rm bath}}}\label{a0_def}.
\end{eqnarray}
Assuming that the displacement of the piston is small $x/X_{\rm ini} \equiv (X-X_{\rm ini})/X_{\rm ini} \ll 1$, the average of Eq. (\ref{eom}) is written as
\begin{eqnarray}
\frac{dV}{dt} &=& -\frac{P_{\rm out}A}{M}\frac{x}{X_{\rm ini}}- \bar{\gamma}V \label{eom_av}
\end{eqnarray}
where we have introduced the viscous friction coefficients $\bar{\gamma}\equiv (\gamma_{\rm gas} + a_0P_{\rm out}A)/M$ and $\gamma_{\rm gas} \equiv 4(1+e)P_{\rm out}A\sqrt{{m}/{2\pi T_{\rm out}}}$. 
The right-hand side of Eq. (\ref{eom_av}) is equivalent to the force acting on a harmonic oscillator in a viscous medium. If the viscous drag is sufficiently small, i.e. ${\epsilon} \to 0$, the motion of the piston is the damped-oscillating type (Fig. \ref{evo}(a)), while the motion turns out to be the over-damped type, if ${\epsilon}$ is not small (Fig. \ref{evo}(b)). 

\subsection{Moderately dense case}\label{dense_time_ev}

Let us examine the validity of SMF for a moderately dense gas. We adopt $d_{\rm in}/\sqrt{A} = 0.1$ which corresponds to $\Phi = 0.105$ at $t = 0$. In Fig. \ref{time_ev_temp}, simulation results for MD, SMF, and the SMF without heat conduction are plotted. It is obvious that the heat conduction plays an important role for the moderately dense gas in contrast to the dilute case (See the inset of Fig. \ref{time_ev_temp}). Although the time evolution of MD for small $t/t_0$ is well predicted by SMF (See the inset of Fig. \ref{time_ev_temp}), the agreement is relatively poor for $0.1<t/t_0 < 0.5$. The agreement for $1.6<t/t_0<2.0$ is also not good, though the difference is not large. Note that the discrepancy for $1.6<t/t_0<2.0$ is not relevant for the efficiency at MP, because we need only $Q_{\rm H}$. The improvement of SMF for $0.1<t/t_0 < 0.5$ is left as a future work.

\begin{figure}[th]
\begin{center}
\includegraphics[scale = 1.6]{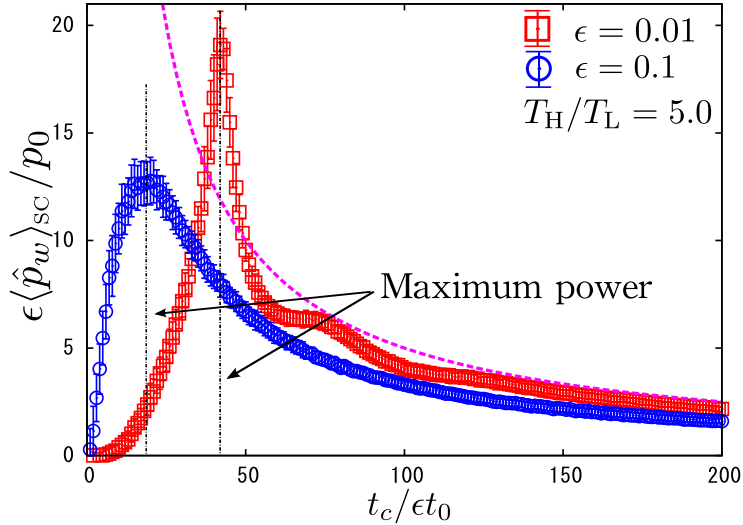}
\caption{(Color online) The average power is plotted against $t_c$. Apparently, there exists $t_c$ for the maximum power operation, which corresponds to the necessary time for gas to expand toward the mechanical equilibrium. The dotted curve drawn as the guide line proportional to $1/t_c$.}
\label{mp}
\end{center}
\end{figure}

\begin{figure*}[th]
\begin{center}
\includegraphics[scale = 1.5]{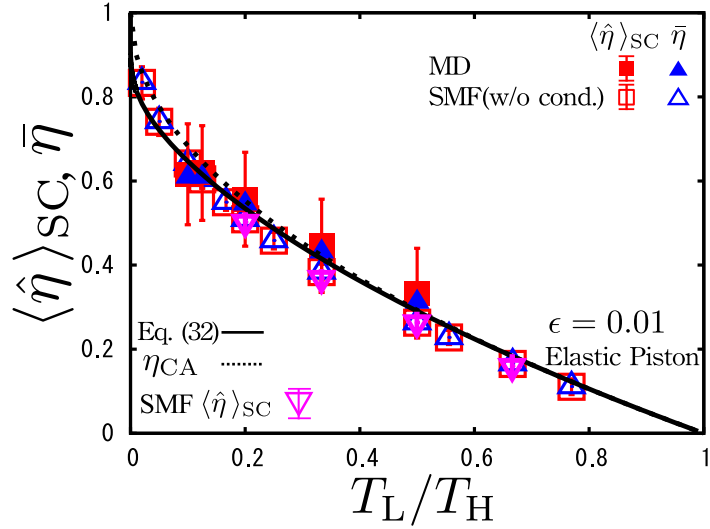}
\caption{(Color online) Efficiencies at maximum power operations for dilute gases for $\epsilon = 0.01$. We plot the result of SMF (open piles). The open squares $\langle\hat{\eta}\rangle_{\rm SC}$ and open triangles $\bar{\eta}$ are simulation data for the SMF without heat conduction, while filled ones are the data for the corresponding MD simulation. The observed efficiencies are close to $\eta_{\rm CA}$ (dashed line) and Eq. (\ref{eta_ana})(solid line). }
\label{eff_mp}
\end{center}
\end{figure*}

\section{Existence of Maximum Power and its Efficiency}\label{sec_exmp}

In this section, we discuss the efficiency of the engine at MP. We show that the efficiency at MP for the dilute gas corresponds to the CNCA efficiency if the piston is sufficiently massive and elastic in Sec. \ref{dilte_eff}, while that for the moderately dense gas is smaller than the CNCA efficiency as will be presented in Sec. \ref{dense_eff}. 

\subsection{Dilute case}\label{dilte_eff}
Let us illustrate that the MP exists for our engine.
We define the work ${\hat{W}_{\rm tot}}$ and the heat spent per a cycle $\hat{Q}_{\rm H}$ as
\begin{eqnarray}
{\hat{W}_{\rm tot}} &\equiv& \oint \frac{1+e}{2}(\hat{P}_{\rm in} - P_{\rm out})Ad\hat{X}\label{def_work_eq},\\
\hat{Q}_{\rm H} &\equiv& \int_{T_{\rm H}}d\hat{Q}_{0}\label{def_qh},
\end{eqnarray}
where $\oint$ and $\int_{T_{\mu}}$ represent the integral over a single cycle and the integral for the bath at $T_{\rm bath} = T_{\mu} (\mu = {\rm H\ or\ L})$, respectively, with the definition in Eq. (\ref{qwall}). It should be noted that Eq. (\ref{def_qh}) is consistent with previous works \cite{izumida,izumida_onsager} and the validity for the definition of work Eq. (\ref{def_work_eq}) is discussed in App. \ref{def_work}. The efficiency for a single operation protocol \cite{CV2014} is defined as
\begin{equation}
\hat{\eta} \equiv \frac{{\hat{W}_{\rm tot}}}{\hat{Q}_{\rm H}}.
\end{equation}
We also introduce the conventional efficiency, which is defined as
\begin{equation}
\bar{\eta} \equiv \frac{\langle{\hat{W}_{\rm tot}}\rangle_{\rm SC}}{\langle\hat{Q}_{\rm H}\rangle_{\rm SC}}.
\end{equation}
In this section, we average the data from 11th cycle to 110th cycle.

The contact time dependence of the power $\hat{p}_w \equiv {\hat{W}_{\rm tot}}/2t_{\rm c}$, for the under-damped type $\epsilon = 0.01$ (squares) and the over-damped type $\epsilon = 0.1$ (circle) are shown in Fig. \ref{mp}, where $T_{\rm H}/T_{\rm L} = 5.0$ and $e = 1.0$ are fixed and $p_0 \equiv  T_{\rm out}/t_0$. Apparently,  the MP is achieved at time $t_c ^{\rm MP}$, which corresponds to the necessary time for the gas to expand toward the mechanical equilibrium. We note that the long time heating or cooling ruins the power, because the extracted work is, at most, $N(T_{\rm H} - T_{\rm L}) {\rm ln} (T_{\rm H}/T_{\rm L})$. Thus, the power decreases as a function of $t_c$: $\langle \hat{p}_w\rangle_{\rm SC} \propto 1/t_c$ for $t_c \gg t_c ^{\rm MP}$, which is drawn as a dashed line in Fig. \ref{mp}. 

We, here, explain that the obtained work is balanced with the work done by the viscous friction for gases. 
Multiplying $V$ onto Eq. (\ref{eom_av}) and integrating over the cycle, we obtain $W_{\rm tot} = \oint M\bar{\gamma}V dX > 0$, because the integral of the left hand side of Eq. (\ref{eom_av}) is zero. Thus, the obtained work is balanced with the work done by the viscous friction for gases.

We present the results for the efficiency at MP (Fig. \ref{eff_mp}) for massive elastic piston $\epsilon = 0.01$ and $e = 1.0$. We discuss the effect of piston mass and its inelasticity in App. \ref{mass_inel}. The open squares $\langle\hat{\eta}\rangle_{\rm SC}$ and triangles $\bar{\eta}$ are the simulation data for the SMF without heat conduction characterized by Eq. (\ref{energy_cons}), while filled ones are the data for the corresponding MD simulation. Although $\bar{\eta}$ and $\langle \hat{\eta}\rangle_{\rm SC}$ are different quantities, they agree with each other. As a comparison with previous studies, we plot the CNCA efficiency $\eta_{\rm CA}$ (dotted lines). 
Our SMF model correctly predicts the efficiency at MP for MD simulations for $\epsilon = 0.01$. We note that the efficiency for our model are close to the CNCA efficiency.

\begin{figure}[h]
\begin{center}
\includegraphics[scale = 0.7]{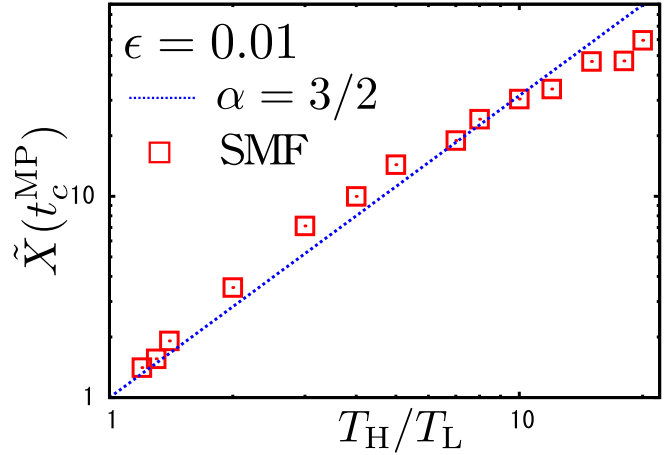}
\caption{(Color online) The volume change of the enclosed gas at MP $\tilde{X}(t_c ^{\rm MP})$ is plotted against $T_{\rm H}/T_{\rm L}$ for $\epsilon = 0.01$.}
\label{xtild}
\end{center}
\end{figure}

Here, we derive the semi-analytical expression on $\bar{\eta}$ on the basis of SMF in the limit $\epsilon \to 0$. In this limit, $\hat{T}_{\rm in}$ rapidly relaxes to bath temperature, right after $T_{\rm bath}$ is switched. The average of the work Eq. (\ref{def_work_eq}) can be approximated by
\begin{eqnarray}
\langle {\hat{W}_{\rm tot}}\rangle_{\rm SC} \simeq N(T_{\rm H} - T_{\rm L}) {\rm ln} \tilde{X}(t_c)\label{work_mp},
\end{eqnarray}
where we have introduced the volume change of the gas through the cycle 
\begin{eqnarray}
\tilde{X}(t_c) \equiv \frac{\langle\hat{X}(t_c)\rangle_{\rm SC}}{\langle\hat{X}(0)\rangle_{\rm SC}}
\end{eqnarray}
and choose $e = 1$. Integrating the equation of the energy conservation (\ref{energy_cons}), we obtain
\begin{eqnarray}
\Delta \hat{U}&=&\hat{Q}_{\rm H} + \hat{E}_{\rm pis} ^{(H)}\label{cons_h}
\end{eqnarray}
where we have introduced $\Delta \hat{U} = 3N(T_{\rm H} - T_{\rm L})/2$ and $\hat{E}_{\rm pis} ^{(H)} \equiv\int_{T_{\rm H}} d\hat{E}_{\rm pis}$.
Averaging Eq. (\ref{cons_h}) and expanding in terms of $\epsilon$, we obtain
\begin{eqnarray}
\langle \hat{Q}_{\rm H}\rangle_{\rm SC} = \frac{3}{2}N(T_{\rm H}-T_{\rm L}) + NT_{\rm H}{\rm ln}\tilde{X}(t_c)+O(\epsilon), \label{qh_mp}
\end{eqnarray}
where we have ignored the heat leak due to the fluctuation of the piston $O(\epsilon)$.
Therefore, the efficiency $\bar{\eta}$ is given by
\begin{eqnarray}
\bar{\eta} &=& \frac{T_{\rm H} - T_{\rm L}}{T_{\rm H} + \frac{3}{2}\frac{T_{\rm H} - T_{\rm L}}{{\rm ln}\tilde{X}(t_c)}}=\frac{\eta_{C}}{1 + \frac{3}{2}\frac{\eta_C}{{\rm ln}\tilde{X}(t_{\rm c})}}\label{eta_ana_0}.
\end{eqnarray}
Assuming that $\tilde{X}(t_{\rm c} ^{\rm MP})$ depends on the power of $T_{\rm H}/T_{\rm L}$ with a power index $\alpha$:
\begin{equation}
\tilde{X}(t_{\rm c} ^{\rm MP}) = \left(\frac{T_{\rm H}}{T_{\rm L}}\right)^{\alpha} = \left(1-\eta_C\right)^{-\alpha},
\end{equation}
we obtain the analytical expression on $\bar{\eta}$ for MP:
\begin{eqnarray}
\bar{\eta}_{\rm MP} &=& \eta_C \left(1-\frac{3}{2\alpha}\frac{\eta_C}{{\rm ln}(1-\eta_C)} \right)^{-1}\nonumber\\
&=& \frac{1}{1 + \frac{3}{2\alpha}}\eta_C + \frac{3}{4\alpha}\left(\frac{1}{1 + \frac{3}{2\alpha}}\right)^2\eta_C ^2+ \frac{\alpha + 6}{8\alpha^2}\left(\frac{1}{1 + \frac{3}{2\alpha}}\right)^3\eta_C ^3+ O(\eta_C ^4)\label{eta_ana},
\end{eqnarray}
which is shown in Fig. \ref{eff_mp} by solid lines. 
The exponent $\alpha$ is estimated from the simulation of SMF, where $\alpha = 1.5$ for $\epsilon = 0.01$ (Fig. \ref{xtild}).The physical meaning of $\alpha$ would be explained in Sec. \ref{sec_ons}. As is shown in Fig. \ref{eff_mp}, Eq. (\ref{eta_ana}) agrees with the results of MD for $\epsilon = 0.01$.
We expect that the exponent $\alpha$ is reduced to $\alpha = 3/2$ in the limit $\epsilon\to 0$ and $e\to 1$, as follows. Although there exists the tiny heat leak during the expansion process, we may approximately ignore the leak because the heating process is almost isochoric, as will be discussed in Sec. \ref{diss_sec}. Recalling Poisson's relation for an adiabatic process of ideal monoatomic gases between state $1$ and $2$: $(T^{(2)} _{\rm in}/T^{(1)} _{\rm in})^{3/2}(X^{(2)} /X^{(1)})= 1$, where $X^{(a)}$ and $T_{\rm in} ^{(a)}$ ($a = 1, 2$) respectively represent the position of the piston and temperature for the state $a$, the exponent $\alpha = 3/2$ agrees with the simulation result. In Sec. \ref{sec_ons}, we will prove that $\alpha = 3/2$ corresponds to the tight coupling condition for the Onsager matrix in linearly irreversible thermodynamics. Substituting the obtained $\alpha = 3/2$ for $\epsilon = 0.01$ into Eq. (\ref{eta_ana}), we obtain
\begin{eqnarray}
\bar{\eta}_{\rm MP} &=& \frac{\eta_C}{2} + \frac{\eta_C ^2}{8} + \frac{5\eta_C ^3}{96} + O (\eta_C ^4)\label{eff_ep001}
\end{eqnarray}
We note that Eq. (\ref{eff_ep001}) is identical to the expansion of $\eta_{\rm CA}$ up to $O (\eta_C ^2)$:
\begin{eqnarray}
\eta_{\rm CA} = \frac{\eta_C}{2} + \frac{\eta_C ^2}{8} + \frac{\eta_C ^3}{16} + O (\eta_C ^4).
\end{eqnarray}
We can here conclude that the efficiency at MP for an engine with an elastic passive piston whose mass is sufficiently massive confining dilute gases is the CNCA efficiency. 

\subsection{Moderately dense case}\label{dense_eff}

\begin{figure}[h]
\begin{center}
\includegraphics[scale = 0.8]{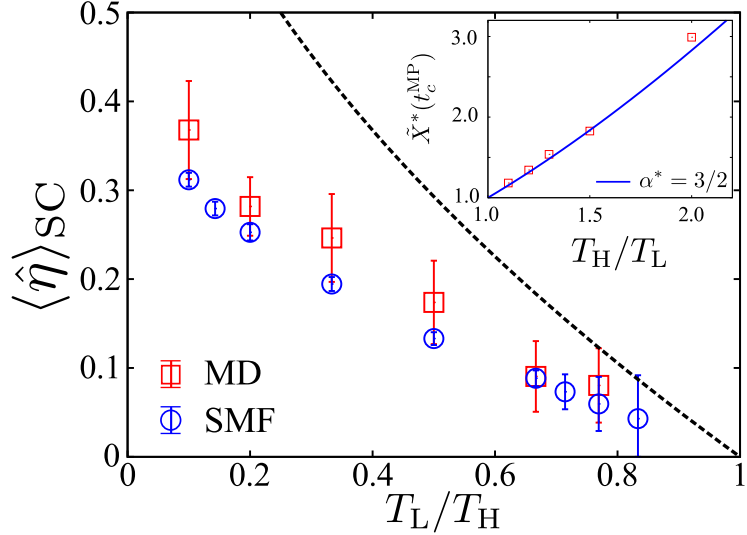}
\caption{(Color online) The main figure represents the efficiency at MP for moderately dense hard core gases. SMF almost correctly predicts the efficiency for MD simulation. We note that the efficiency is much smaller than the CNCA one, which is caused by the inevitable heat flux $d{Q}_{J}$. The inset represents the expansion ratio $\tilde{X}^*$ defined in Eq. (\ref{xtilde_modify}) for moderately dense gases. The exponent $\alpha$ is estimated to be $\alpha^* \simeq 3/2$.}
\label{eff_mp_modify}
\end{center}
\end{figure}

We have analyzed the efficiency for dilute gases in the previous subsection. Here, we discuss the efficiency at MP for a moderately dense hard core gas.
The efficiency at MP is plotted in the main figure of Fig. \ref{eff_mp_modify}, where SMF model almost correctly predicts the results of our MD simulation. The data for SMF at $T_{\rm H}/T_{\rm L} = 1.2, 1.3, 1.4$ are averaged over $1.0\times10^4$ cycles after $10$ cycles for initial relaxation to improve their numerical accuracy. The other data are averaged from 11th cycle to 110th cycle. We find that the efficiency for moderately dense hard core gases is smaller than that for dilute ones to compensate the heat flux $J_{\rm in}$ as will be shown in the next section.

\section{Linearly irreversible thermodynamics}\label{sec_ons}

In the previous section, we have suggested that the efficiency at MP output for the dilute gas can be described by the CNCA efficiency in the limit $\epsilon\to 0$ and $e\to 1$, while that for the moderately dense gas is smaller than the CNCA efficiency. In this section, we show that results in linear non-equilibrium situation $\eta_C\to 0$ can be understood by the relations between the currents ${\mathcal J}_i$ and the thermodynamic forces ${\mathcal X}_i$ on the basis of the Curie-Prigogine symmetry principle \cite{onsager}:
\begin{eqnarray}
{\mathcal J}_1 &=& L_{11}{\mathcal X}_1 + L_{12}{\mathcal X}_2\label{j1_matrix},\\
{\mathcal J}_2 &=& L_{21}{\mathcal X}_1 + L_{22}{\mathcal X}_2\label{j2_matrix},
\end{eqnarray}
where the Onsager matrix satisfies $L_{11},L_{22} \geq 0, L_{12} = L_{21}$ and $\det L_{ij} = L_{11}L_{22} - L_{12}L_{21} \geq 0$. In the following, we assume that the piston is elastic $e = 1.0$ and massive limit $\epsilon\to 0$, and we abbreviate the average of an arbitrary stochastic quantity $\hat{\mathcal A}$ as ${\mathcal A} = \langle \hat{\mathcal A}\rangle_{\rm SC}$. We examine the dilute gas in Sec. \ref{ons_dilute} and clarify the finite density effect in Sec. \ref{ons_dense}.

\subsection{Dilute case}\label{ons_dilute}

Let us derive the Onsager matrix $L_{ij}$ in our setup for the dilute gas following Refs. \cite{izumida_onsager,izumida_nonlinear}. We consider the linear non-equilibrium situation as $T_{\rm H, L} = T \pm \Delta T/2$, where $T$ and $\Delta T$ are the mid-temperature $T \equiv (T_{\rm H} + T_{\rm L})/2$ and the temperature difference $\Delta T = T_{\rm H}-T_{\rm L}$, respectively, satisfying $\Delta T/T \ll 1$. Here, the total entropy production per a unit cycle $\Delta\sigma = -Q_{\rm H}/T_{\rm H} -Q_{\rm L}/T_{\rm L}$ is rewritten as
\begin{eqnarray}
\Delta\sigma &=& -\frac{W_{\rm tot}}{T} + \frac{\Delta T}{T^2}Q_{\rm H}\label{epro_dilute},
\end{eqnarray}
where we have used $W_{\rm tot} = Q_{\rm H} + Q_{\rm L}$ and $\Delta T/T \ll 1$. On the basis of the relation
\begin{eqnarray}
\frac{\Delta\sigma}{2t_c} &=& {\mathcal J}_1{\mathcal X}_1 + {\mathcal J}_2{\mathcal X}_2,
\end{eqnarray}
 ${\mathcal J}_i$ and ${\mathcal X}_i$ are respectively given by
\begin{eqnarray}
{\mathcal J}_1 &=& \frac{T}{2t_c},\ {\mathcal J}_2 = \frac{Q_{\rm H}}{2t_c}\label{j_12},\\
{\mathcal X}_1 &=& -\frac{W_{\rm tot}}{T^2},\ {\mathcal X}_2 = \frac{\Delta T}{T^2} = \frac{\eta_C}{T}\label{x_12}. 
\end{eqnarray}
Let us derive $L_{11}$ and $L_{21}$ by taking $\eta_C = \Delta T/T \to 0$. $W_{\rm tot}$ is written as
\begin{eqnarray}
W_{\rm tot} \simeq N\eta_CT{\rm ln}\tilde{X}(t_c) - 2a_0 NT \int_{X_{\rm L}} ^{X_{\rm H}}V\frac{dX}{X}\label{work_ons}.
\end{eqnarray}
The first term on the right-hand side of Eq. (\ref{work_ons}) vanishes in the limit $\eta_C\to 0$. Then, from Eqs. (\ref{j1_matrix}), (\ref{j_12}), and (\ref{x_12}) we obtain
\begin{eqnarray}
L_{11} &=& \frac{T^2}{4t_c N}\frac{1}{\tilde{E}}\geq 0\label{L11},\\
\tilde{E} &\equiv& \int_{X_{\rm L}} ^{X_{\rm H}}a_0V\frac{dX}{X}.
\end{eqnarray}
Here, we have introduced $\tilde{E}$ as the inevitable dissipation due to the finite velocity of the piston. Now the heat $Q_{\rm H}$ is given by
\begin{eqnarray}
Q_{\rm H} &=& \frac{3}{2}N\frac{\Delta T}{T}T + N\left(T+\frac{\Delta T}{2}\right){\rm ln}\tilde{X}(t_c)- N\left(T+\frac{\Delta T}{2}\right)a_0\int_{X_{\rm L}} ^{X_{\rm H}}V\frac{dX}{X},\label{q_h_ons}
\end{eqnarray}
which can be rewritten as
\begin{eqnarray}
\frac{Q_{\rm H}}{2t_c} &=& \frac{T^2}{4t_c}\frac{{\rm ln}\tilde{X}(t_c)- \tilde{E}}{\tilde{E}}\left(-\frac{W_{\rm tot}}{T^2}\right)
\simeq \frac{T^2}{4t_c}\frac{{\rm ln}\tilde{X}(t_c)}{\tilde{E}}{\mathcal X}_1,\\
L_{21} &=& \frac{T^2}{4t_c\tilde{E}}{{\rm ln}\tilde{X}(t_c)}\label{L21},
\end{eqnarray}
in the leading order of $W_{\rm tot}/T$ and the limit $\eta_C \to 0$. From Eq. (\ref{a0_def}), we have used ${\rm ln}\tilde{X}(t_c) \gg \tilde{E} = O(\epsilon)$ in the limit $\epsilon\to 0$.
Next, let us determine $L_{12}$ and $L_{22}$. $L_{12}$ can be determined from the condition $W_{\rm tot} = 0$, i.e., the work-consuming state:
\begin{eqnarray}
W_{\rm tot} = N{\mathcal X}_2T^2{\rm ln}\tilde{X}(t_c) - 2NT\tilde{E}= 0.\label{w_consume}
\end{eqnarray}
Then, we obtain the reciprocal relation
\begin{eqnarray}
L_{12} &=& \frac{T^2}{4t_c\tilde{E}}{{\rm ln}\tilde{X}(t_c)} = L_{21}\label{L12}.
\end{eqnarray}
Taking terms depending only on $\Delta T$ in Eq. (\ref{q_h_ons}), we obtain
\begin{eqnarray}
\frac{Q_{\rm H}}{2t_c} &\simeq& \frac{1}{2t_c}\left(\frac{3}{2}NT^2 + \frac{NT^2}{2}{\rm ln}\tilde{X}(t_c)\right)\frac{\Delta T}{T^2},\\
L_{22} &=& \frac{NT^2}{2t_c}\left(\frac{3}{2} + \frac{1}{2}{\rm ln}\tilde{X}(t_c)\right)\geq 0\label{L22},
\end{eqnarray}
where we have ignored the higher order term including $a_0$. Equations (\ref{L11}), (\ref{L21}), (\ref{L12}) and (\ref{L22}) are the explicit expressions of the Onsager matrix.

Here, we show that $\alpha = 3/2$ corresponds to the tight coupling limit of the Onsager matrix, where flux ${\mathcal J}_1$ is proportional to ${\mathcal J}_2$. Because the determinant is readily calculated as 
\begin{eqnarray}
\det L_{ij} &=& \left(\frac{T^4}{8t_c ^2}\frac{1}{\tilde{E}}\right)\left(\frac{3}{2} + \frac{1}{2}{\rm ln}\tilde{X}(t_c)\right) - \left(\frac{T^2}{4t_c\tilde{E}}{{\rm ln}\tilde{X}(t_c)}\right)^2\nonumber\\
&=& \frac{T^4}{8t_c ^2 \tilde{E}}\left\{\frac{3}{2} + \frac{1}{2}{\rm ln}\tilde{X}(t_c) - \frac{({\rm ln}\tilde{X}(t_c))^2}{2\tilde{E}}\right\}\nonumber\\
&=&\frac{T^4}{8t_c ^2 \tilde{E}}\left(\frac{3}{2} + \frac{1}{2}{\rm ln}\tilde{X}(t_c) - \frac{{\rm ln}\tilde{X}(t_c)}{\eta_C}\right) \simeq\frac{T^4}{8t_c ^2 \tilde{E}}\left(\frac{3}{2}-\alpha \right) \geq 0,\label{lij_ineq_dilute}
\end{eqnarray}
where we have used Eq. (\ref{w_consume}) with Eq. (\ref{x_12}), i.e. ${\rm ln}\tilde{X}(t_c)/2\tilde{E} = 1/\eta_C$ and ${\rm ln}\tilde{X} = -\alpha{\rm ln}(1-\eta_C) \simeq \alpha\eta_C + \alpha\eta_C ^2/2 + O(\eta_C ^3)$ under the nearly equilibrium condition $\eta_C\to 0$.
The tight coupling limit $\det L_{ij} = 0$ corresponds to $\alpha = 3/2$, which is equal to the value obtained in Sec. \ref{sec_exmp}. The CNCA efficiency is derived on the basis of Eqs. (\ref{j1_matrix}) and (\ref{j2_matrix}) in the tight coupling limit, following the similar procedure in Ref. \cite{vdb}. It should be noted that the control parameter for our engine is not ${\mathcal X}_1$ but ${\mathcal J}_1$, in contrast to Ref. \cite{vdb}.

\subsection{Moderately dense case}\label{ons_dense}

We stress that the efficiency at MP of the engine for the moderately dense gas is much smaller than the CNCA efficiency even in linear non-equilibrium regime $\eta_C\ll 1$, which is the result of the inevitable loose coupling of the Onsager matrix $L_{ij} ^*$ as follows. Solving the average of Eq. (\ref{energy_cons_modify}) in terms of $T_{\rm in}$, we obtain
\begin{eqnarray}
T_{\rm in }(t) &=& T_{\rm bath}(1 - a_0 ^*(t) V(t)) + O(\epsilon^2)\\
a_0 ^* (t)&\equiv& \frac{a_0}{1+4\Phi(t) g_0(\Phi(t)) + \tilde{j}_{\rm in}(t)},
\end{eqnarray}
where we have introduced the scaled flux $\tilde{j}_{\rm in} = \{T_{\rm bath}/(T_{\rm bath} - T_{\rm in})\}d{Q}_{J}/dt$. See also Eq. (\ref{a0_def}) for the comparison with the dilute case. Because the additional heat flux $d{Q}_{J}$ exists, Eqs. (\ref{epro_dilute}) and (\ref{q_h_ons}) are, respectively, replaced by
\begin{eqnarray}
\Delta\sigma &=& -\frac{W_{\rm tot}^*}{T} + \frac{\Delta T}{T^2}Q_{\rm H} + \frac{1}{T}{Q}_{J}\label{epro_dense},\\
Q_{\rm H} &=&\frac{3}{2}N\frac{\Delta T}{T}T + N\left(T+\frac{\Delta T}{2}\right){\rm ln}\tilde{X}^*(t_c) + {Q}_{J} ^{\rm H} - N\left(T+\frac{\Delta T}{2}\right)\int_{X_{\rm L}} ^{X_{\rm H}} a_0 ^*(t)V\frac{dX}{X},\nonumber\\ \label{q_h_ons_mod}
\end{eqnarray}
where we have introduced
\begin{eqnarray}
{Q}_{J} &\equiv& \sum_{\mu = {\rm H, L}} {Q}_{J} ^{\mu},\\
{Q}_{J} ^{\mu}&\equiv& -\int_{T_{\mu}} d{Q}_{J} \label{q1_def},\\
W_{\rm tot} ^* &\equiv& N\eta_C T{\rm ln}\tilde{X}^*(t_c) - 2NT \int_{X_{\rm L}} ^{X_{\rm H}}a_0 ^*V\frac{dX}{X},\\
\tilde{X}^*(t_c) &\equiv& \frac{\langle \hat{X}(t_c)\rangle_{\rm SC} - 4v_{\rm ex}/A}{\langle \hat{X}(0)\rangle_{\rm SC}- 4v_{\rm ex}/A}.\label{xtilde_modify}
\end{eqnarray}
Note that the sign of ${Q}_{J} ^{\rm H}$ and ${Q}_{J} ^{\rm L}$ are positive and negative respectively, and they are $O(\Delta T)$, while ${Q}_{J} > 0$ is $O(\Delta T ^2)$ (See Eqs. (\ref{qwall1}), (\ref{jin_modify}), and (\ref{q1_def})). We have taken into account the effect of the finite excluded volume $v_{\rm ex} \equiv N\pi d_{\rm in}^3/6$ up to $O(\Phi)$ for $\tilde{X}^*$, where we have approximated Eq. (\ref{eos_hard}) as $P_{\rm in} \simeq n_{\rm in}T_{\rm in}(1+4\Phi) \simeq n_{\rm in}T_{\rm in}/(1-4\Phi)$. Following the similar procedure in Sec. \ref{ons_dilute}, we obtain the Onsager matrix ${\mathcal J}^* _i = \sum_j L_{ij}^* {\mathcal X}^* _j$ with $i,j = 1,2$ as
\begin{eqnarray}
L_{11} ^* &\equiv& \frac{T^2}{4t_c N}\frac{1}{\tilde{E} ^*}\geq 0\label{L11_st},\\
L_{21} ^* &\equiv& \frac{T^2}{4t_c\tilde{E}^*}{{\rm ln}\tilde{X}^*(t_c)} = L_{12} ^*\label{L21_st},\\
L_{22} ^* &\equiv& \frac{NT^2}{2t_c}\left(\frac{3}{2} + \frac{1}{2}{\rm ln}\tilde{X}^*(t_c) + \tilde{q}\right) \geq 0\label{L22_st},
\end{eqnarray}
where we have introduced $\tilde{E}^* \equiv  \int_{X_{\rm L}} ^{X_{\rm H}}(V/X) a_0 ^*(t){dX}$ and $\tilde{q} \equiv {Q}_{J} ^{\rm H}/N\Delta T + {Q}_{J} T/N\Delta T ^2 > 0$. Note that ${\mathcal J}^* _1 \equiv T/2t_c$, ${\mathcal J}^*_2 \equiv (Q_H  + TQ_J/\Delta T)/2t_c$, ${\mathcal X}^* _1 \equiv - W_{\rm tot} ^*/T$, and ${\mathcal X}^* _2 \equiv \eta_C /T$ have been introduced. We have checked that a positive current $\tilde{q}$ exists even if $T_{\rm H}\sim T_{\rm L}$ as $\tilde{q} \simeq 1.91$ for the operation of MP with $T_{\rm H}/T_{\rm L} = 1.1$ through the simulation of SMF. 

Let us derive the value of $\alpha^*$ for the tight coupling condition: ${\rm det}L_{ij}^* = 0$. Introducing $\tilde{X}^* = (T_{\rm H}/T_{\rm L})^{\alpha^*}$ with the aid of the parallel argument to derive Eq. (\ref{lij_ineq_dilute}), the tight coupling condition for $L_{ij} ^*$ is reduced to
\begin{eqnarray}
\det L_{ij} ^*&=& \left(\frac{T^4}{8t_c ^2}\frac{1}{\tilde{E}^*}\right)\left(\frac{3}{2} + \frac{1}{2}{\rm ln}\tilde{X}^*(t_c) + \tilde{q}\right) - \left(\frac{T^2}{4t_c\tilde{E}^*}{{\rm ln}\tilde{X}^*(t_c)}\right)^2\nonumber\\
&\simeq& \frac{T^4}{8t_c ^2 \tilde{E}^*}\left(\frac{3}{2}-\alpha^* +\tilde{q}\right) = 0.
\end{eqnarray}
Thus, we obtain $\alpha^*$ for the tight coupling condition as
\begin{eqnarray}
\alpha^* = \frac{3}{2} + \tilde{q}.
\end{eqnarray}
However, this condition cannot be satisfied if the finite positive current $\tilde{q}$ exists as observed in our simulation, because we find that $\alpha^* \simeq 3/2$ holds through our simulation (inset of Fig. \ref{eff_mp_modify}). Thus, we conclude that the tight coupling condition for moderately dense gases is not satisfied because of $\tilde{q}$. The loose coupling property of the Onsager matrix can be rewritten as the heat leak from the hot heat reservoir into the cold hot reservoir: $J_{\rm leak} = {\mathcal J}_2 ^*- (L_{22} ^*{\mathcal J}_1 ^*/L_{11} ^*)$ \cite{izumida_nonlinear}. From our relations (\ref{L11_st})-(\ref{L22_st}), the heat leak is expressed as
\begin{eqnarray}
J_{\rm leak} = \frac{{\rm det}L_{ij} ^*}{L_{11}^*}{\mathcal X}_2 ^* = \frac{NT}{2t_c}\eta_C \left(\frac{3}{2} + \tilde{q} - \alpha^*\right) \simeq \frac{NT}{2t_c}\eta_C \tilde{q}> 0.
\end{eqnarray} 

As mentioned in Sec. \ref{dilte_eff}, the exponent $\alpha=3/2$ for dilute gases is that for adiabatic processes. Therefore, we can examine whether such idea can be used even in moderately dense gases. As is well-known, Poisson's relation for a moderately dense gas can be written as:
\begin{eqnarray}
\left(\frac{T^{(2)} _{\rm in}}{T^{(1)} _{\rm in}}\right)^{3/2}\left(\frac{X^{(2)} - 4v_{\rm ex}/A}{X^{(1)}- 4v_{\rm ex}/A}\right)= 1.
\end{eqnarray}
Therefore, we also have the relation $\tilde{X}^* \simeq (T_{\rm H}/T_{\rm L})^{3/2}$, i.e. $\alpha^* = 3/2$ for quasi-static adiabatic processes.
Although this agreement may be accidental because the heat leak exists in the process, it is interesting to look for the reason why Poisson's relation works
well.

\section{Discussion}\label{diss_sec}
\begin{figure}[th]
\begin{center}
\includegraphics[scale = 1.6]{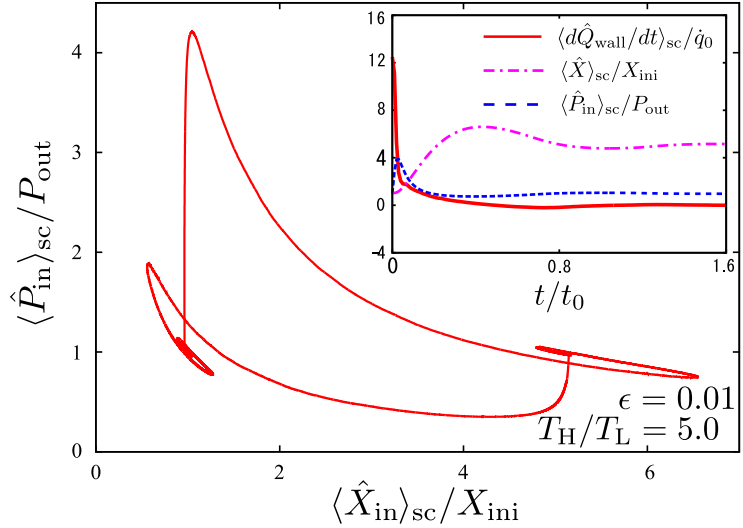}
\caption{(Color online) The main figure represents the pressure-volume figure for $\epsilon = 0.01, T_{\rm H}/T_{\rm L} = 5.0$. The inset represents the time evolution of the heat flux from the thermal wall (solid line), the position of the piston (chain line) and the pressure for the enclosed gas (dashed line) for $0<t/t_0<1.6$. The heat process ends fast, and can be regarded as an isochoric one. }
\label{pv_fig}
\end{center}
\end{figure}

Let us discuss the difference between our results and previous results. 
Here, we explain that our engine contains isochoric and quasi-adiabatic heating/cooling processes, i.e., our engine is similar but different from the Otto engine. 
The pressure-volume graph for $\epsilon = 0.01$ and $T_{\rm H}/T_{\rm L} = 5.0$ is plotted in the main figure of Fig. \ref{pv_fig}. 
We also plot the time evolutions of the heat flux (solid line), the piston position (chain line), and the pressure (dashed line) for $0<t/t_0<1.6$ in the inset of Fig. \ref{pv_fig}, where the heat flux is scaled by $\dot{q}_0 \equiv 5.0\times 10^5 T_{\rm out}/t_0$. 
We notice that the heating process ends readily at $t/t_0 \sim 0.1$. 
Then, the system expands with smaller heat flux which is less than 10\% of the isochoric regime for $t/t_0<0.5$. For $0.5<t/t_0<1.6$, the system is
almost adiabatic, i.e. the heat flux is negligible. 
Thus, our engine is similar but different from the Otto engine. As we can see in the inset of Fig. \ref{pv_fig}, the piston moves only for $t/t_0 > 0.1$, which might be related to the reason why we can use Poisson's equation for the adiabatic expansion in our analysis.

Let us explain the reason why the heat flux $d{Q}_{J}$ for a moderately dense gas is relevant to the efficiency at MP in contrast to the conventional finite time thermodynamics. As a counter example, let us consider the finite time Carnot cycle, which contains isothermal and adiabatic processes. 
When we attach the thermal bath to the gas, the amount of heat flux for a finite time Carnot cycle is too small and $d{Q}_{J}$ does not exist, because the temperature of the gas and that of the bath are essentially identical as the result of the adiabatic processes with mechanical control of the piston. On the other hand, the amount of heat flux in our engine is large because the temperature of the gas and that of the bath are different when we attach the bath onto the gas. Thus, the effect of the heat flux $d{Q}_{J}$ is significant for the efficiency for an engine with a passive piston.

For a macroscopic piston in the limit $\epsilon \to 0$, the one-dimensional momentum transfer model (Eqs. (\ref{rule1}) and (\ref{rule2})) is too simple for the realistic motion of the piston, where the side-wall friction \cite{sano_hayakawa,ref_fri}, the excitation of atoms on the piston surface \cite{itami} and tilting of the piston,  etc. should be relevant for the real piston motion. In App. \ref{friction_effect}, we discuss the effect of side-wall friction on the efficiency for our protocol and show that the side-wall friction lowers the efficiency.

The model considered in this paper might be unrealistic if the gas is regarded as a molecular gas, because the mass of the piston must be much larger
than the mass of each molecule and adhesion between molecules and walls cannot be ignored in such a small engine. Our model, however, would be experimentally realized through two kinds of setups: colloidal suspensions with a semi-permeable membrane and a highly excited granular gas with a movable piston. Although the hydrodynamic interaction between colloids is important, the osmotic pressure between two dilute solutions separated by a semi-permeable membrane is described by van't Hoff's formula which has an identical form to the state equation for ideal gases. Similarly, inhomogeneity and non-Gaussianity of granular gases can be suppressed, at least, for a specific setup of a highly agitated granular gas  \cite{andrea}. Thus, our model can be regarded as a simplified and idealized one for such systems. We also note that our result is expected to be basically valid even in thermodynamic limit, though this paper only discusses small systems which contains only 200 particles.

\section{Concluding Remarks}\label{sec_con}
In this paper, we have investigated the efficiency at MP for an engine with a passive piston. 
We have considered an operation protocol for a hard core gas partitioned by a massive piston (Figs. \ref{setup} and \ref{protocol}).
SMF has been proposed and its relevance has been demonstrated from the comparison of its results with those of the MD simulation for both dilute gas (Fig. \ref{evo}) and the moderately dense gas (Fig. \ref{time_ev_temp}). 
We have found the existence of the MP in Fig. \ref{mp} and examined the efficiency at MP for the dilute gas in Fig. \ref{eff_mp}. The efficiency at MP for dilute gases is close to the CNCA efficiency for an elastic and massive piston. We have derived the analytic expressions for the efficiency at MP on the basis of SMF as Eqs. (\ref{eta_ana}) and (\ref{eff_ep001}). To understand the linear non-equilibrium regime, we have derived the Onsager matrix explicitly Eqs. (\ref{L11}), (\ref{L21}), (\ref{L12}), and (\ref{L22}), and have found that the tight coupling condition is satisfied for the dilute gas. In contrast to the dilute gas, we have found that the efficiency at MP for moderately dense gases is smaller than the CNCA efficiency even for an elastic and massive piston in linear non-equilibrium regime (Fig. \ref{eff_mp_modify}). We have clarified the importance of the heat flux when $T_{\rm bath}$ is switched, which induces the inevitable loose coupling for the Onsager matrix.

To improve SMF model, we need to solve hydrodynamic equations under the moving boundary in contrast to the treatment in this paper. We also need to investigate the nonlinear Onsager matrix to understand the efficiency in nonlinear non-equilibrium regime \cite{izumida_nonlinear, esposito_2010}. Finally, because thermodynamic studies of engines without any force controls are little known so far,  their experimental studies will be expected near future.

\section*{Acknowledgement}
We are grateful for useful discussion with Y. Izumida, K. Kanazawa, A. Puglisi, L. Cerino, S. Ito, E. Iyoda, and T. Sagawa. This work is supported by the Grants-in-Aid for Japan Society for Promotion of Science (JSPS) Fellows (Grants No. 26$\cdot$2906), and JSPS KAKENHI (Grant Nos. 25287098). This work is also partially supported by the JSPS core-to-core program for Nonequilibrium dynamics for soft matter and information.

\appendix
\section{Derivation of Eqs. (\ref{q_wall_pwall1}), (\ref{qwall}), and (\ref{qwall1})}\label{derive_qwall}
It is known that VDF for a hard core gas under the heat flux $\hat{J}_{\rm in}$ \cite{garzo,resibor} is given by
\begin{eqnarray}
\phi_{\rm flux}(\boldmath v) &=& \left(1 + v_xc({\boldmath v})\hat{J}_{\rm in} \right) \phi_0({\boldmath v})\label{vdf_mod},
\end{eqnarray}
where we have introduced 
\begin{eqnarray}
\phi_0(\boldmath v) &\equiv& \prod_{\mu = x,y,z}\phi_0(v_{\mu}, \hat{T}_{\rm in}).
\end{eqnarray}
In Eq. (\ref{vdf_mod}), $c(\boldmath v)$ is written as
\begin{eqnarray}
c({\boldmath v}) &\equiv& - \frac{4}{5\hat{n}_{\rm in}\hat{T}_{\rm in}}\left(\frac{mv^2}{2\hat{T}_{\rm in}}-\frac{5}{2}\right).
\end{eqnarray}
The energy flows $d\hat{Q}_{0}/dt$ and $d\hat{Q}_{J}/dt$ can be calculated as follows. The heat flows outgoing $\hat{q}^{\rm out} _{\rm wall}$ and incoming $\hat{q}^{\rm in} _{\rm wall}$ through the wall are, respectively, given by
\begin{eqnarray}
\hat{q}^{\rm out} _{\rm wall} &=& \left\{\int_{-\infty} ^{\infty}dv_ydv_z\int_{-\infty} ^0 dv_x \frac{m{\boldmath v}^2}{2}(-v_x)\hat{n}_{\rm in}A\phi_{\rm flux}({\boldmath v})\right\}\label{qout}\\
\hat{q}^{\rm in} _{\rm wall} &=&  \left\{\int_{-\infty} ^{\infty}dv_ydv_z\int_{-\infty} ^0 dv_x(-v_x)\hat{n}_{\rm in}A\phi_{\rm flux}({\boldmath v})\right\} \left\{\int_{-\infty} ^{\infty}dv_ydv_z\int_{0} ^{\infty} dv_x \frac{m{\boldmath v}^2}{2}\phi_{\rm wall}({\boldmath v}, T_{\rm bath})\right\}.\nonumber\\\label{qin}
\end{eqnarray}
Substituting Eqs. (\ref{qout}) and (\ref{qin}) into $d\hat{Q}_{\rm wall}= (\hat{q}^{\rm in} _{\rm wall} - \hat{q}^{\rm out} _{\rm wall})dt$, we obtain Eqs. (\ref{q_wall_pwall1}), (\ref{qwall}), and (\ref{qwall1}).

\section{Time evolution for the profile of the temperature}\label{grad_temp_md}
\begin{figure*}[h]
\centering
\includegraphics[scale = 1.2]{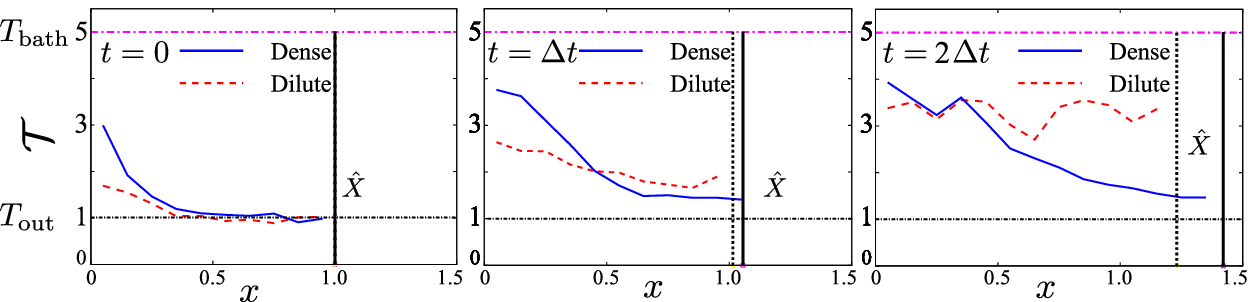}
\caption{(Color online) The profiles of the temperature with the time interval $\Delta t =  5.0\times10^{-3}\sqrt{MA/T_{\rm out}}$ right after we change $T_{\rm bath}$. The solid and dotted curves represent the profile of the temperature for dense $d_{\rm in}/\sqrt{A}=0.1$ and dilute $d_{\rm in}/\sqrt{A} = 0.01$ gases, respectively. The vertical solid and dotted lines represent the position of the piston enclosing dense and dilute gases, respectively.}
\label{prof_temp_fig}
\end{figure*}

In this appendix, we show that the time evolution for the temperature profile strongly depends on the density of the enclosed gas. In Fig. \ref{prof_temp_fig}, we plot the profiles of the temperature with the time interval $\Delta t = 5.0\times10^{-3}\sqrt{MA/T_{\rm out}}$ right after we change $T_{\rm bath}$. Here, the solid and the dotted curves represent the profile of the temperature for dense $d_{\rm in}/\sqrt{A}=0.1$ and dilute $d_{\rm in}/\sqrt{A} = 0.01$ gases, respectively. The vertical solid and dotted lines represent the position of the piston enclosing dense and dilute gases, respectively. The gradient of the temperature for the dilute gases relaxes much faster than that for the dense gases. As we increase the value of $d_{\rm in}/\sqrt{A}$, the relaxation time for the gradient becomes larger, which can be captured by introducing $d{Q}_{J}$ as in the text.

\section{On the definition of work}\label{def_work}
In the text, we define the work as ``{Pressure $\times$ Volume change}," which is not trivial. In this appendix, we justify the definition, i.e. we decompose the change of the kinetic energy of piston into heat and work by considering the path probability of $(\hat{X}(t),\hat{V}(t))$ under $\hat{T}_{\rm in}(t) = T_{\rm in}$. The discussion here is the extension of Ref. \cite{itami2} toward the case that the volume of the enclosed gas fluctuates in time. Let us consider the path probability for the forward evolution ${\mathcal P}([\hat{X},\hat{V}|\tau)$ of $(\hat{X},\hat{V})$ during the interval $\tau$ from $(\hat{X}(0), \hat{V}(0))$ to $(\hat{X}(\tau), \hat{V}(\tau))$ and the backward one ${\mathcal P}([\hat{X},\hat{V}]^{\dagger}|\tau)$ from $(\hat{X}(\tau), -\hat{V}(\tau))$ to $(\hat{X}(0), -\hat{V}(0))$, where $n$ collisions between the piston and particles take place at time $\{t_i\}_{i=1} ^n$ with $0 = t_0 < t_1 < \cdots < t_n = \tau$. The jump rates for the piston velocity from $V_{i-1} \equiv \hat{V}(t_{i-1})$ to $V_{i} \equiv \hat{V}(t_{i})$ at the piston position $X_{i-1} \equiv \hat{X}(t_{i-1})$ caused by collisions from particles inside and outside the container are, respectively, written as
\begin{eqnarray}
{\mathcal W}_{\rm in}(V_{i} \leftarrow V_{i-1}|X_{i-1}) &\equiv& n_{\rm in}(X_{i-1})A \int_{-\infty} ^{\infty} dv |v- {V}_{i-1}|\Theta(v- {V}_{i-1}) \phi({v}, {{T}_{\rm in}})\nonumber\\
&&\times\delta\left(V_i - V_{i-1} -\frac{P_v(V_{i-1})}{M}\right),\\
{\mathcal W}_{\rm out}(V_{i} \leftarrow V_{i-1}) &\equiv& n_{\rm out}A \int_{-\infty} ^{\infty} dv |v - V_{i-1}|\Theta(V_{i-1} -v) \phi({v}, {T_{\rm out}})\nonumber\\
&&\times\delta\left(V_i - V_{i-1} -\frac{P_v(V_{i-1})}{M}\right),\\
{\mathcal W}_{\rm tot}(V_{i} \leftarrow V_{i-1}|X_{i-1}) &\equiv& {\mathcal W}_{\rm in}(V_{i} \leftarrow V_{i-1}|X_{i-1})+ {\mathcal W}_{\rm out}(V_{i} \leftarrow V_{i-1}).
\end{eqnarray}
The escape rate per a unit time $\kappa(V_{i-1}|X_{i-1})$ for $(X_{i-1},V_{i-1})$ is represented as
\begin{eqnarray}
\kappa(V_{i-1}|X_{i-1}) &=& \int_{-\infty} ^{\infty}{dV'} W_{\rm tot}(V' \leftarrow V_{i-1}|X_{i-1})\nonumber\\
&=& n_{\rm in}(X_{i-1})A \int_{V_{i-1}} ^{\infty}|v-V_{i-1}|\phi_0(v, {T_{\rm in}})dv\nonumber\\
&&+  n_{\rm out}A \int_{-\infty} ^{V_{i-1}}|v-V_{i-1}|\phi_0(v, {T_{\rm out}})dv,
\end{eqnarray}
Thus, ${\mathcal P}([X,V]|\tau)$ and ${\mathcal P}([X,V]^{\dagger}|{\tau})$ are represented as
\begin{eqnarray}
{\mathcal P}([X,V]|{\tau}) &=& \exp\left[-\sum_{i = 0} ^{n-1}\int_{t_i} ^{t_{i+1}} \kappa(V_i|X({s_i}))ds_i \right]\left[\prod_{i=1} ^{n}{\mathcal W}_{\rm tot}(V_{i} \leftarrow V_{i-1}|X_{i-1})\right], \\
{\mathcal P}([X,V]^{\dagger}|{\tau}) &=& \exp\left[-\sum_{i = 0} ^{n-1}\int_{t_i} ^{t_{i+1}} \kappa(-V_i|X({s_i}))ds_i \right]\left[\prod_{i=1} ^{n}{\mathcal W}_{\rm tot}(-V_{i-1} \leftarrow -V_{i}|X_{i-1})\right].\nonumber\\
\end{eqnarray}
Here, the position of the piston at time $t_i<s_i<t_{i+1}$ is given by $X({s_i}) \equiv X_i + V_i(s_i - t_i)$. 
We obtain
\begin{eqnarray}
\int_{t_i} ^{t_{i+1}} \left\{\kappa(V_i|X_{s_i}) - \kappa(-V_i|X_{s_i})\right\}ds_i &=& -N \ln\left(\frac{X_{i+1}}{X_i}\right) + n_{\rm out}AV_i(t_{i+1} - t_i)\nonumber\\
&=&-\beta_{\rm in}\int_{X_i} ^{X_{i+1}}n_{\rm in}(X)T_{\rm in}AdX+ \beta_{\rm out}{P_{\rm out}A}V_j(t_{i+1} - t_i),\nonumber\\ \label{kratio}
\end{eqnarray}
\begin{eqnarray}
\ln\left\{\frac{{\mathcal W}_{\rm tot}(V' \leftarrow V|X)}{{\mathcal W}_{\rm tot}(-V \leftarrow -V'|X)}\right\} = \left\{ \begin{array}{l}
\displaystyle \beta_{\rm in}\frac{m(v'^{2} - v^2) }{2}\equiv\beta_{\rm in}\Delta E_{\rm in}(V'> V)\\ \\
\displaystyle\beta_{\rm out}\frac{m(v'^{2} - v^2) }{2}\equiv\beta_{\rm out}\Delta E_{\rm out} (V' < V),\label{wratio}\\
\end{array} \right.
\end{eqnarray}
Here we have introduced the inverse temperature $\beta_{\nu}\equiv1/T_{\nu}$ and the energy change of $\nu$ side gas $\Delta E_{\nu}$ through the piston fluctuation $(\nu = {\rm in,\ out})$. Using Eqs. (\ref{kratio}) and (\ref{wratio}), we obtain the following expression on the definition of the work:
\begin{eqnarray}
\ln\left\{\frac{{\mathcal P}([X,V]|{\tau})}{{\mathcal P}([X,V]^{\dagger}|{\tau})}\right\} &=& \beta_{\rm in}{\Delta Q_{\rm in}} +\beta_{\rm out}{\Delta Q_{\rm out}} + \Delta S_{\rm inel},\\
\Delta E_{\rm in} &=& \Delta{Q_{\rm in}} -\int_{X_{\rm ini}} ^{X_{\tau}} \frac{1+e}{2}n_{\rm in}T_{\rm in}AdX,\label{in_decomp}\\
\Delta E_{\rm out} &=& \Delta{Q_{\rm out}} + \frac{1+e}{2}P_{\rm out}A\int^{X_{\tau}}_{X_{\rm ini}}dX,\\
\Delta S_{\rm inel} &\equiv& \frac{1-e}{2}\int_{X_{\rm ini}} ^{X_{\tau}} \left\{n_{\rm in}T_{\rm in} - P_{\rm out}\right\}AdX
\end{eqnarray}
where we have introduced the abbreviation $V_0 \equiv \hat{V}(0), X_{\tau} \equiv \hat{X}(\tau)$ and $V_{\tau} \equiv \hat{V}(\tau)$. From Eq. (\ref{in_decomp}), the change of the internal energy for the enclosed gas $\Delta E_{\rm in}$ is apparently decomposed into the change of work and heat. Thus, we adopt the definition of work Eq. (\ref{def_work_eq}) in the text. For force-controlled engines, we usually define their works using only $P_{\rm out}$. However, we define the work using the pressure difference, because our engine is not force-controlled.

\section{Mass and inelasticity of piston}\label{mass_inel}
\begin{figure*}[th]
\begin{center}
\includegraphics[scale = 1.2]{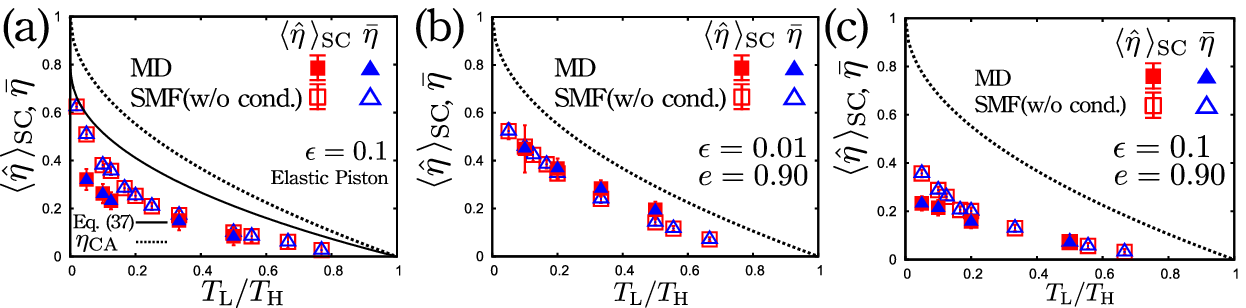}
\caption{(Color online) Efficiencies at maximum power operations for dilute gases for (a) $e = 1.0$ and $\epsilon = 0.1$, (b) $e = 0.9$ and $\epsilon = 0.01$, and (c) $e = 0.9$ and $\epsilon = 0.1$. The open squares $\langle\hat{\eta}\rangle_{\rm SC}$ and open triangles $\bar{\eta}$ are simulation data for the SMF without heat conduction. The observed efficiencies for light and inelastic pistons are much smaller than $\eta_{\rm CA}$ (dashed line). We also plot Eq. (\ref{eta_ana}) as a solid line in $(a)$ which overestimates the simulation results.}
\label{eff_mp_light}
\end{center}
\end{figure*}

In this appendix, the effects of mass and inelasticity of the piston are studied. Similar to Sec. \ref{sec_exmp}, we find the existence of maximum power for light or inelastic pistons. We plot the efficiency at MP for (a) $e = 1.0$ and $\epsilon = 0.1$, (b) $e = 0.9$ and $\epsilon = 0.01$, and (c) $e = 0.9$ and $\epsilon = 0.1$ in Fig. \ref{eff_mp_light}. The observed efficiencies for light and inelastic pistons are much smaller than $\eta_{\rm CA}$ (dashed line). Through our simulation, we find $\alpha = 0.79$ for $e = 1.0$ and $\epsilon = 0.1$.
We plot Eq. (\ref{eta_ana}) in $(a)$, while the observed efficiencies are also smaller than Eq. (\ref{eta_ana}). The existence of $\epsilon$ lowers the efficiency from $\eta_{\rm CA}$ even at the leading order $O(\eta_C)$, because $\alpha = 0.79 < 3/2$ for $\epsilon = 0.1$. The higher order correction for $\epsilon$ would be necessary for better agreement.

\section{Effect of side-wall friction}\label{friction_effect}
\begin{figure*}[!bottom]
\centering
\includegraphics[scale = 1.1]{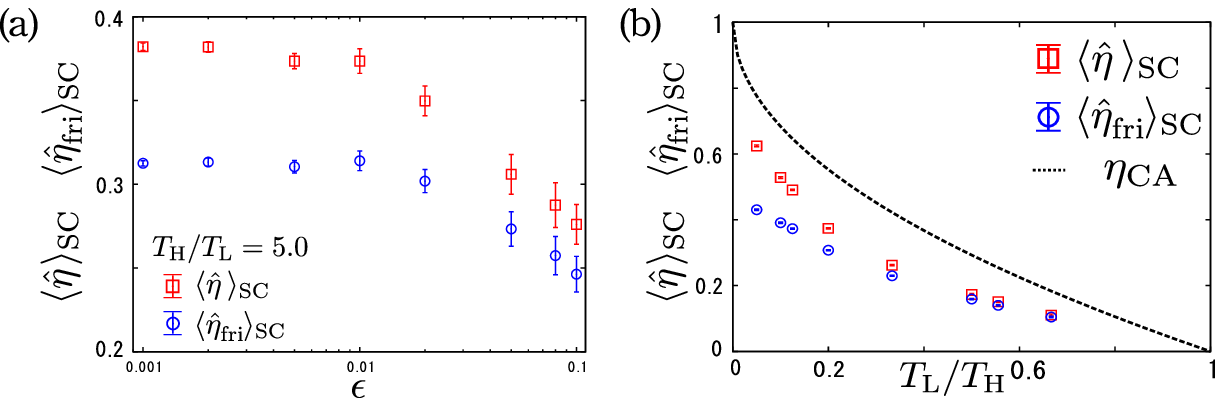}
\caption{(Color online) The efficiency at MP under side-wall friction. The asymptotic behavior of the efficiencies in $\epsilon \to 0$ limit (a), and their temperature dependence for $\epsilon = 0.001$ and $\gamma/\gamma_{\rm gas} = 2.0$ (b). The friction on the sidewall lowers the efficiency.}
\label{eff_fri}
\end{figure*}

In this appendix, we discuss the effect of the side-wall friction on the efficiency for an engine with a passive piston, which exists for realistic situations. We implement the linear friction on the side-wall as $\hat{F}_{\rm fri} = - \gamma \hat{V}$. Then, the equation of motion Eq. (\ref{eom}) turns out to be
\begin{equation}
M\frac{d\hat{V}}{dt} = \hat{F}_{\rm in} +\hat{F}_{\rm out} + \hat{F}_{\rm fri}\label{eom_fri}
\end{equation}
We assume that $\gamma$ does not depend on $\epsilon$ and $\gamma/\gamma_{\rm gas} = O(1)$, where the motion of the piston becomes the over-damped type, even if the piston is heavy. Because the side-wall friction can be regarded as that attached with a zero temperature bath, we define the efficiency under friction \cite{ref_fri} by introducing the frictional heat:
\begin{eqnarray}
\hat{Q}_{\rm fri} &\equiv& \oint \gamma \hat{V}^2 dt\\
\hat{\eta}_{\rm fri} &\equiv& \frac{{\hat{W}_{\rm tot}}}{\hat{Q}_{\rm H} + \hat{Q}_{\rm fri}}
\end{eqnarray}
The simulated data for the efficiency at MP with $\gamma / \gamma_{\rm gas}=2.0$ and $e = 1.0$ are plotted in Fig. \ref{eff_fri}. The asymptotic behavior of $\langle\hat{\eta}\rangle_{\rm SC}$ and $\langle\hat{\eta}_{\rm fri}\rangle_{\rm SC}$ in the limit $\epsilon \to 0$ for $T_{\rm H}/T_{\rm L} = 5.0$ are shown in Fig. \ref{eff_fri} (a). In Fig. \ref{eff_fri} (b), we plot the temperature dependence of $\langle\hat{\eta}\rangle_{\rm SC}$ and $\langle\hat{\eta}_{\rm fri}\rangle_{\rm SC}$ at MP with $\epsilon = 0.001$, where the efficiencies are lower than $\eta_{\rm CA}$(see Fig. \ref{eff_mp} (a)). Thus, as expected, the friction on the sidewall lowers the efficiency.


\end{document}